\documentclass[12pt]{article}
\usepackage{latexsym,axodraw}
\def\dslash{\raisebox{1pt}{$\slash$} \hspace{-7pt} \partial}
\def\Aslash{\hspace{3pt}\raisebox{1pt}{$\slash$} \hspace{-9pt} A}
\def\Dslash{\hspace{3pt}\raisebox{1pt}{$\slash$} \hspace{-9pt} D}
\def\bea{\begin{eqnarray}}
\def\eea{\end{eqnarray}}
\def\be{\begin{equation}}
\def\ee{\end{equation}}
\def\nn{\nonumber}
\def\a{& \hspace{-7pt}}

\def\Z{{\bf Z}}
\begin{document}

\thispagestyle{empty}

\begin{center}
\hfill CERN-TH/2001-267\\
\hfill ROMA-1325/01 \\
\hfill SISSA-77/2001/EP \\

\begin{center}

\vspace{1.7cm}

{\LARGE\bf Anomalies in orbifold field theories}

\end{center}

\vspace{1.4cm}

{\sc C.A. Scrucca$^{a}$, M. Serone$^{b}$,
L. Silvestrini$^{c}$ and F. Zwirner$^{c}$}\\

\vspace{1.2cm}

${}^a$
{\em CERN, CH-1211 Geneva 23, Switzerland}
\vspace{.3cm}

${}^b$
{\em ISAS-SISSA, Via Beirut 2-4, I-34013 Trieste, Italy} \\
{\em INFN, sez. di Trieste, Italy}
\vspace{.3cm}

${}^c$
{\em Dip. di Fisica, Univ. di Roma ``La Sapienza'' and INFN, sez. di Roma} \\
{\em P.le Aldo Moro 2, I-00185, Rome, Italy}
\end{center}

\vspace{0.8cm}

\centerline{\bf Abstract}
\vspace{2 mm}
\begin{quote}\small
We study the constraints on models with extra dimensions arising 
from local anomaly cancellation. We consider a five-dimensional 
field theory with a $U(1)$ gauge field and a charged fermion, 
compactified on the orbifold $S^1/(\Z_2 \times \Z_2^\prime)$. We show 
that, even if the orbifold projections remove both fermionic 
zero modes, there are gauge anomalies localized at the fixed 
points. Anomalies naively cancel after integration over the 
fifth dimension, but gauge invariance is broken, spoiling 
the consistency of the theory. We discuss their implications 
for realistic supersymmetric models with a single Higgs 
hypermultiplet in the bulk, and possible cancellation 
mechanisms in non-minimal models.
\end{quote}

\vfill

\newpage
\setcounter{equation}{0}

\section{Introduction}

Theories formulated in $D>4$ space-time dimensions may lead 
to a geometrical understanding of the problems of mass 
generation and symmetry breaking. Orbifold compactifications 
\cite{Dixon} of higher-dimensional theories are simple and 
efficient mechanisms to reduce their symmetries and to generate 
four-dimensional (4-D) chirality. Phenomenologically interesting 
orbifold models can be formulated, either as explicit string 
constructions or as effective higher-dimensional field theories.

The field-theoretical approach to orbifolds is currently fashionable 
because of its apparent simplicity and flexibility. However, 
it is well known that the rules for the construction of consistent
string-theory orbifolds are quite stringent, and automatically 
implement a number of consistency conditions in the corresponding
effective field theories: in particular, the cancellation of gauge, 
gravitational and mixed anomalies. Since anomalies are infrared 
phenomena, if we start from a consistent string model (`top--down' 
approach), anomaly cancellation must find an appropriate description
in the effective field theory. Such a description, however, may 
be non-trivial, as for the Green--Schwarz \cite{gs} or the inflow 
\cite{ch} mechanisms. If, instead, we decide to work directly at 
the field-theory level (`bottom--up' approach), great care is 
needed, since orbifold projections do not necessarily preserve 
the quantum consistency of a field theory (as discussed, for example, 
in \cite{hw}). In particular, the question of anomaly cancellation must be
explicitly addressed. 

A first step in this direction was taken in ref.~\cite{Cohen}, 
which discussed the chiral anomaly in a five-dimensional (5-D)
theory compactified on the orbifold $S^1/\Z_2$. 
It was found that, in such a simple context, naive 4-D anomaly 
cancellation is sufficient to ensure 5-D 
anomaly cancellation. For a 5-D fermion of unit charge, and a chiral
action of the $\Z_2$ projection, the 5-D anomaly 
is localized at the orbifold fixed points, and is proportional 
to the 4-D anomaly:
\be
\label{prevres}
\partial_M J^M (x,y) = {1 \over 4} \,\Big[
\delta(y) + \delta (y- \pi R) \Big] \, {\cal Q}(x,y) \,,
\ee
where\footnote{We work on the orbifold covering space $S^1$, and 
we normalize the $\delta$-functions so that, for $y_0 \in [0,2 
\pi R )$ and $0< \epsilon < 2 \pi R - y_0$, 
$\int_{- \epsilon}^{2 \pi R - \epsilon}
d y \delta(y - y_0) f(y) = f(y_0)$.} $J^M$ is the 5-D 
current and
\be
\label{qcal}
{\cal Q}(x,y) = \frac {g_5^2}{16 \pi^2}\,
F_{\mu\nu}(x,y)\, \tilde F^{\mu\nu}(x,y)
\ee
is proportional to the 4-D chiral anomaly from a charged Dirac spinor
in the external gauge potential $A_{\mu}(x,y)$. 
In our notation: $M=[(\mu=0,1,2,3),4]$;  $x\equiv (x^{0,1,2,3})$ are the 
first four coordinates, $y \equiv x^4$ is the fifth 
coordinate, compactified on a circle of radius $R$; $y=0,\pi R$ are the 
two fixed points with respect to the $Z_2$ symmetry $y \rightarrow - y$;
$g_5$ is the 5-D gauge coupling constant. 

In this letter we show that the phenomenon discussed in
\cite{Cohen} does not persist in more general cases. 
To be definite, we consider a 5-D field theory 
with a $U(1)$ gauge field $A_M$ and a massless fermion
$\psi$ of unit charge, compactified on the orbifold 
$S^1/(\Z_2 \times \Z_2^\prime)$. The action of the 
two parities are $y \rightarrow -y$ and $y' \rightarrow
- y'$, respectively, where $y' = y - \pi R /2$. Both
the gauge and the fermion fields are taken to be periodic
on the circle. We decompose the Dirac spinor $\psi$ into
left and right spinors with parities $(+,-)$ and $(-,+)$,
respectively: $\psi \equiv \psi^{+-} + \psi^{-+}$. 
Notice that a standard fermion mass term is forbidden
by the $\Z_2 \times \Z_2^\prime$ symmetry. As for the gauge field, 
we assign $(+,+)$ parities to $A_\mu$, $(-,-)$ to $A_4$.
Although the theory has no massless 4-D chiral fermion, 
a non-vanishing anomaly is induced, given by eq.~(\ref{ano}) 
below.

The theory can be trivially supersymmetrized, by embedding 
its field content in a $U(1)$ vector multiplet and a 
charged hypermultiplet. From the point of view of 
anomalies, our simple example reproduces the essential
features of a recently proposed phenomenological model
\cite{Barber}, whose light spectrum contains just the states 
of the Standard Model (SM), with an anomaly-free fermion
content. The underlying 5-D theory is supersymmetric, with 
vector multiplets containing the SM gauge bosons, and 
hypermultiplets containing the SM quarks and leptons. 
In addition, the model of ref.~\cite{Barber} has just one 
charged hypermultiplet, which contains the SM Higgs boson. 
Such a model has received some attention because it may give 
a prediction for the Higgs mass, even though it was recently 
shown \cite{Nilles} that the Higgs self-energy receives a 
quadratically divergent one-loop contribution. The latter
corresponds to the appearance of a Fayet--Iliopoulos 
(FI) term, with divergences localized at the orbifold 
fixed points, which immediately hints at a possible connection 
with anomalies.

The content of the present letter is organized as follows.
We begin by showing that, even if there are no 4-D
massless fermions in the spectrum, our simple 5-D theory is
actually anomalous. Localized anomalies, with opposite signs, 
appear at the fixed points of the two orbifold projections. 
The integrated anomaly vanishes, reflecting the absence of 
any one-loop anomaly among 4-D massless states, but there 
are anomalous triangle diagrams when at least one of the 
external states is a massive Kaluza--Klein (KK) mode. We focus 
our attention on the $U(1)^3$ gauge anomaly, which we explicitly 
compute along the lines of \cite{Cohen}. In realistic extensions,
such as \cite{Barber},
similar results would hold for the $U(1)_Y^3$, $U(1)_Y$--$SU(2)^2_L$ 
and $U(1)_Y$--gravitational anomalies. We then argue that this anomaly 
leads to a breakdown of 4-D gauge invariance.
Hence, in its minimal form, the model is 
inconsistent, even as an effective low-energy theory. Next,
we consider the supersymmetric extension of our simple theory,
and compute the precise expression for the one-loop FI term. 
Finally, we discuss the possible modifications that could 
restore the consistency of the theory. 

\section{$U(1)$ anomalies}
\label{sec:anomalies}

In this section, we take the theory defined in the Introduction
and we compute the $U(1)^3$ anomaly, following closely the method 
and the notation of ref.~\cite{Cohen} (for an early computation of 
this type, see also ref.~\cite{Binetruy}).

The KK wavefunctions $\xi^{ab}$ for fields $\varphi^{ab}$ 
of definite $\Z_2 \times \Z_2^\prime$ parities $(a,b=\pm)$ are defined as:
\begin{eqnarray}
\xi^{++}_{n \ge 0}(y) \a\equiv\a \frac{\eta_n}{\sqrt{\pi R}} 
\cos \frac {2 n y}R \, , \;\;\;\;\;
\xi^{+-}_{n>0}(y) \equiv \frac{1}{\sqrt{\pi R}} 
\cos \frac {(2 n -1)y}R \, , \nn \\ 
\xi^{--}_{n>0}(y) \a\equiv\a \frac{1}{\sqrt{\pi R}} 
\sin \frac {2 n y}R \, , \;\;\;\;\;
\xi^{-+}_{n>0}(y) \equiv \frac{1}{\sqrt{\pi R}} 
\sin \frac {(2 n -1)y}R \,,
\label{eq:icchese} 
\end{eqnarray}
where $\eta_n$ is $1/\sqrt{2}$ for $n=0$ and $1$ for $n>0$. They form 
a complete orthonormal basis of periodic functions on $S^1$, with given 
$\Z_2 \times \Z_2^\prime$ parities. 
The Fourier modes of a field $\varphi^{ab}$ are defined as:
\begin{eqnarray}
\varphi^{ab}_{n}(x) \a \equiv \a\int_0^{2 \pi R} \!\!\! dy\,  
\xi^{ab}_n(y) \, \varphi^{ab}(x,y) \,,
\label{eq:fthiggs}  
\end{eqnarray}
and have a mass given by $m_{2n+(ab-1)/2}$, where $m_n = n/R$. 

In the gauge $A_4 = 0$, the 4-D Lagrangian for the Fourier modes 
$\psi_n \equiv \psi_n^{+-} + \psi_n^{-+}$ can be written as 
\be
\label{eq:lagrangian}
{\mathcal L}= \sum_{m,n} 
\bar \psi_m \Big[(i\,\dslash - m_{2n-1}) \delta_{mn} 
- g_5 \, \Aslash_{mn} \Big] \psi_n \,,
\ee
where $\Aslash_{mn} \equiv \Aslash_{mn}^{+-} P_+ + \Aslash_{mn}^{-+} P_-$, 
with $P_\pm = (1 \pm \gamma_5)/2$, and~\footnote{Notice that the 
$A_{\mu\,mn}^{\pm\mp}$ are not Fourier modes of the type (\ref{eq:fthiggs}),
but can be easily related to them. One finds:
$A^{\mu \pm \mp}_{mn} = (\eta_{|m-n|}^{-1} A^\mu_{|m-n|} \pm 
\eta_{|m+n-1|}^{-1} A^\mu_{|m+n-1|})/\sqrt{\pi R}$.}
\be
A^{\pm \mp}_{\mu\,mn}(x) \equiv \int_0^{2 \pi R} \!\!\! dy\, 
\xi^{\pm \mp}_m(y) \, \xi^{\pm \mp}_n(y) \, A_\mu(x,y)
\ee
in terms of the $U(1)$ connection $A_\mu$.

Interpreting $\psi_n$ as a single fermion with a flavour index and chiral 
couplings to the gauge field $A^{\mu}_{mn}$ through the currents
${J^{\mu}_{\pm}}_{mn} = \bar \psi_m \gamma^\mu P_\pm \psi_n$,
it is straightforward to adapt the standard computation of anomalies 
to obtain:
\be
\partial_\mu {J^\mu_{\pm}}_{mn} = \pm \Big(m_{2m-1} J^4_{\pm mn} 
+ m_{2n-1} J^4_{\mp mn}\Big) \pm \frac {g_5^2}{32 \pi^2}\, \sum_{k>0}
F_{\mu\nu mk}^{\pm \mp}\, \tilde F^{\mu\nu \pm \mp}_{kn} \,,
\label{mm}
\ee
where ${J^4_{\pm}}_{mn} = \bar \psi_m i\gamma_5P_\pm \psi_n$.
Equation~(\ref{mm}) can be easily Fourier-transformed back to configuration 
space by convolution with $\xi^{\pm \mp}_m(y)\, \xi^{\pm \mp}_n(y)$.
Using completeness, this yields
\be
\partial_M J^{M}_\pm(x,y) =
\pm \frac 12 \sum_{k>0} [\xi_k^{\pm \mp}(y)]^2 \,{\cal Q}(x,y) \,,
\ee
where the quantity ${\cal Q}$ was defined in eq.~(\ref{qcal}). 
The anomaly in the vector current $J^{M}(x,y) = J^M_+(x,y) + J^M_-(x,y)$ 
is then proportional to
\be
\sum_{k > 0} \Big[(\xi_k^{+-}(y))^2 - (\xi_k^{-+}(y))^2\Big] =
\frac 14\, e^{-2\,i\,y/R} \sum_{l=-\infty}^\infty \delta(y - l \pi R/2) \,,
\ee
hence
\be
\partial_M J^{M} (x,y) = \frac 18
\Big[\delta(y) - \delta(y - \pi R/2) 
+ \delta(y - \pi R) - \delta(y - 3\pi R/2) \Big] {\cal Q}(x,y) \,.
\label{ano}
\ee
Therefore, although the integrated anomaly vanishes, there are anomalies, 
localized at the fixed points, that are equal in magnitude to $1/4$
(or $1/2$ if we sum the contribution from identified fixed points) 
of the anomaly from a 4-D Weyl fermion. The full 5-D theory 
is thus inconsistent (at least in its minimal form). 

Let us now rewrite eq.~(\ref{ano}) in terms of standard Fourier 
modes of the current and gauge fields. Recalling that both have $(+,+)$ 
parities, the Fourier transform of (\ref{ano}) takes the form:
\be
q_M J^M_n(q) = \frac 1{g_5} \sum_{i,j=0}^\infty \int \! \frac {d^4p}{(2\pi)^4}\,
q_M \,T^{M \alpha \beta}_{n i j}(p,q)\, A_{\alpha i}(p) \,A_{\beta j}(q-p) \,,
\ee
where
\be
q_M T^{M \alpha \beta}_{n i j}(p,q) = \frac {g_4^3}{2\sqrt{2}\,\pi^2}\,
\eta_n \, \eta_i \, \eta_j \, \delta_{n+i+j,odd} \, 
\epsilon^{\alpha \beta \mu \nu} p_\mu \, q_\nu \,,
\label{qT}
\ee
with $g_4 = g_5/\sqrt{2 \pi R}$.
This quantity encodes the triangular anomaly between three external 
KK modes of the photon with indices $(n,i,j)$, as illustrated in 
fig.~1.

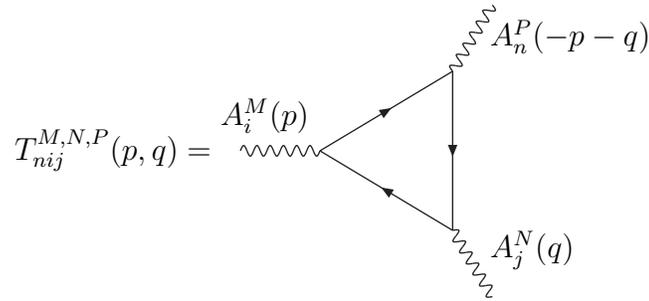
\begin{figure}[h]
\begin{center} 
\begin{picture}(280,130)(0,-10)
\put(35,47){$T_{nij}^{M,N,P}(p,q) =$}
\put(113,60){$A^M_i(p)$}
\put(215,10){$A^N_j(q)$}
\put(215,90){$A^P_n(-p-q)$}
\ArrowLine(150,50)(200,80)
\ArrowLine(200,80)(200,20)
\ArrowLine(200,20)(150,50)
\Photon(120,50)(150,50){2}{6}
\Photon(200,80)(215,105){2}{6}
\Photon(200,20)(215,-5){2}{6}
\end{picture}
\caption{The 1-loop anomalous diagram.}
\vskip -10pt
\end{center}
\end{figure}

This anomaly vanishes for $n+i+j=$ even, and in particular for
$n = i = j = 0$, reflecting the fact that there is no 4-D anomaly
for the massless modes:
all non-vanishing anomalous diagrams involve at least one massive mode. 
These diagrams make the full theory inconsistent. However, it may be asked 
whether the low-energy effective theory obtained by integrating out all 
massive modes could be consistent. This is not the 
case, because gluing such diagrams through heavy lines produces 4-D gauge 
symmetry breaking effective interactions among zero-modes.
Consider for instance a 3-loop diagram obtained by gluing two anomalous 
triangles through two massive photons, as depicted in fig.~2. 
This represents a contribution to the two-point function $\Pi^{\mu \nu}$ 
of the zero-mode photon that violates gauge invariance. 
The non-vanishing longitudinal component of $\Pi^{\mu \nu}$ is encoded 
in $q_\mu q_\nu \Pi^{\mu \nu}(q)$, which feels only the anomalous part
of the triangular subdiagram \cite{agwgj}. Another example is the
four-point function involving two longitudinal and two transverse
zero-mode photons, which receives a non-vanishing finite two-loop
contribution controlled by the anomaly.

\begin{figure}[h]
\begin{center} 
\begin{picture}(270,90)(0,13)
\put(0,47){$\Pi^{\mu \nu}(q) =$}
\put(60,60){$A^\mu_0(q)$}
\put(240,60){$A^\nu_0(-q)$}
\ArrowLine(100,50)(150,80)
\ArrowLine(150,80)(150,20)
\ArrowLine(150,20)(100,50)
\Photon(70,50)(100,50){2}{6}
\Photon(150,80)(180,80){2}{6}
\Photon(150,20)(180,20){2}{6}
\Photon(230,50)(260,50){2}{6}
\ArrowLine(180,80)(230,50)
\ArrowLine(230,50)(180,20)
\ArrowLine(180,20)(180,80)
\end{picture}
\caption{The 3-loop anomalous contribution to the photon two-point function.}
\vskip -10pt
\end{center}
\end{figure}
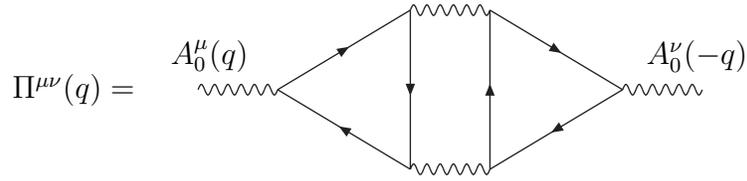

These gauge anomalies could be computed in an independent way by using 
their well-known relation with chiral anomalies and index theorems, which 
is particularly clear in Fujikawa's approach. 
In this formalism, the integrated chiral anomaly of a 4-D Dirac fermion is 
encoded
in the quantity ${\rm Tr}_{D=4}\,[\gamma_5] = {\rm index}(\Dslash)$, where
$\Dslash$ is the Dirac operator. In the case of a 5-D theory compactified 
on $S^1$, the anomaly vanishes, since the Hilbert space splits into two 
identical components of opposite chirality and 
${\rm Tr}_{D=5}\,[\gamma_5] = 0$. This can be easily extended to an 
$S^1/(\Z_2 \times \Z_2^\prime)$ orbifold compactification. 
The trace must now be 
restricted to invariant states only; this can be achieved by inserting into 
the unconstrained trace a $\Z_2 \times \Z_2^\prime$ projector $P$. Denoting
by $g$ and $g^\prime$ the generators of $\Z_2$ and $\Z_2^\prime$ respectively,
the explicit expression of this projector is 
$P=\frac 14(1+g+g^\prime + g g^\prime)$. Each element in $P$, when inserted 
in the trace, leads to a so-called equivariant index of the Dirac operator. 
This has a non-vanishing support only at the fixed points of the element.
The identity in 
$P$ gives a vanishing result as in the $S^1$ compactification. Similarly, 
the $g g^\prime$ element also gives a vanishing contribution, because it
generates a translation along the compact direction that does not affect 
chirality. On the other hand, the elements $g$ and $g^\prime$ act chirally
on the Dirac fermion $\psi$ ($g \psi g^{-1} = \gamma_5 \psi$, 
$g^\prime \psi {g^\prime}^{-1} = -\gamma_5 \psi$) 
and give a non-vanishing contribution. Both have two fixed points,  
and the integrated anomaly is thus :
\bea
{\rm Tr}_{D=5}\,[P\,\gamma_5] \a=\a 
\frac 14 \sum_{i=1}^4 {\rm index}\,(\Dslash)|_{y_i} \\
\a=\a \frac 14 \int d^4x \Big[{\cal Q}(x,0) + {\cal Q}(x,\pi R) 
- {\cal Q}(x,\pi R/2) - {\cal Q}(x,3\pi R/2)\Big] \,, \nn
\eea
where the relative sign between the contributions associated with $g$ and 
$g^\prime$ is due to their opposite action on fermions. This leads to
(\ref{ano}).

\section{Supersymmetric models}

The result found for the anomalies in section \ref{sec:anomalies} can
be trivially extended to supersymmetric configurations, where the
$U(1)$ gauge field belongs to a 5-D $N=1$ vector multiplet and the
Dirac fermion $\psi$ to a charged hypermultiplet. As such, all the
above considerations apply also to the model of ref.
\cite{Barber}. In particular, focusing on the Higgs hypermultiplet,
doublet under $SU(2)_L$ with hypercharge $Y=+1$, we get a localized 
$U(1)_Y^3$ anomaly that is twice the one in (\ref{ano}).

The reader may wonder whether such localized anomaly has any relation 
with the one-loop FI term recently found in \cite{Nilles}. The 
method of the previous section can be easily extended to the 
computation of the full one-loop FI term. The relevant part of 
the Lagrangian is 
\bea
\label{lagfi}
{\mathcal L}\a=\a -\sum_{m,n} (\phi_m^{++})^\dagger
\Big[(\Box + m_{2n}^2)\delta_{mn} - g_5 \, D_{mn}^{++} \Big] \phi_n^{++} \nn \\
\a\;\a -\, \sum_{m,n} (\phi_m^{--})^\dagger \Big[(\Box +
m_{2n}^2)\delta_{mn} + g_5 \, D_{mn}^{--} \Big] \phi_n^{--} \,, 
\eea
where $\phi_m^{\pm\pm}$ are the modes, defined according to eqs. 
(\ref{eq:icchese}) and (\ref{eq:fthiggs}), of the two scalars in 
the Higgs hypermultiplet, and
\be
D^{\pm\pm}_{mn}(x) \equiv \int_0^{2 \pi R} \!\!\! dy\,  
\xi^{\pm\pm}_m(y) \, \xi^{\pm\pm}_n(y) \, D(x,y) \,,
\ee
where $D(x,y)$ is the third component of the triplet of $N=2$ auxiliary 
fields. Considering again the mode indices as flavour indices, we find
for the FI term:
\be
\label{eq:FI-inizio}
{\cal F}(x) = \sum_{n\ge 0} T_n \Big(D_{nn}^{++} - D_{nn}^{--}\Big)(x) \,,
\ee
where
\be
T_n = i\,g_5\,\int \frac {d^4p}{(2 \pi)^4} \frac 1{p^2 - m_{2n}^2} 
= \frac {g_5}{16 \pi^2} \Bigg(\Lambda^2 -  m_{2n}^2 \ln 
\frac{\Lambda^2 + m_{2n}^2}{m_{2n}^2} \Bigg)  \,,
\label{Tn}
\ee
where $\Lambda$ is an ultraviolet cutoff.
By Fourier-transforming back to configuration space, we can write
\be
{\cal F}(x) = \int_0^{2 \pi R} \!dy \,\xi(y) D(x,y) \, ,
\ee 
where the exact profile of $\xi(y)$ can be explicitly evaluated. By 
first summing over the KK states, the 4-D momentum 
integral is convergent for generic $y$, yielding:
\be
\xi(y) = \sum_{n\ge 0}\,T_n\,\Big[(\xi_n^{++}(y))^2 - (\xi_n^{--}(y))^2\Big] = 
\frac {g_5}{8 \pi^5 R^3} \Bigg[\zeta\Big(3,\frac {2 \tilde y}{\pi R}\Big) 
+ \zeta\Big(3,1-\frac {2 \tilde y}{\pi R}\Big) \Bigg] \,,
\label{fiexp}
\ee
where $\tilde y = y - \pi R /2 \sum_{l>0} \theta(y - l\pi R/2)$ is the 
restriction of $y$ to the interval $[0,\pi R/2[$.
Equation~(\ref{fiexp}) diverges as $\tilde y^{-3}$ when $\tilde y$ tends to
$0$, which corresponds to $y$ approaching one of the four fixed points 
$y_i = (i-1)\pi R/2$ $(i=1,2,3,4)$. 
Away from the fixed points, there is only a finite 
bulk contribution. The divergent part of $\xi(y)$ is easily evaluated by
going back to (\ref{Tn}) and using:
\bea
\a\a \sum_{k \ge 0} \Big[(\xi_k^{++}(y))^2 - (\xi_k^{--}(y))^2\Big]
= \frac 14 \sum_{i=1}^4 \delta(y - y_i) \nn \,, \\
\a\a \sum_{k \ge 0} m_{2k}^2 \Big[(\xi_k^{++}(y))^2 - (\xi_k^{--}(y))^2\Big]
= - \frac 1{16} \sum_{i=1}^4 \delta^{\prime\prime}(y-y_i) \,.
\eea
Hence, the structure of the FI term is:
\be
{\cal F}(x) = \frac {g_5}{64 \pi^2} \sum_{i=1}^4 \Big[\Lambda^2 D(x,y_i) 
+ \frac 12 \ln (\Lambda R) \,D^{\prime\prime}(x,y_i) \Big]
+ \int_0^{2 \pi R} \!dy\, K(y)\, D(x,y) \,,
\ee
with $K(y)$ being a finite function. Therefore, the divergent part 
of the induced FI term is localized at the orbifold fixed points, 
as the anomaly. This is a remnant of the relation between FI terms 
and mixed $U(1)$--gravitational anomalies in supersymmetric theories. 

\section{Outlook}

We have seen that orbifold field theories can be anomalous even
in the absence of an anomalous spectrum of zero modes. It is then
important to understand whether there exist anomaly cancellation
mechanisms, and whether they can be consistently implemented: for
definiteness, we discuss this issue by making reference once more 
to the case of $S^1/(\Z_2 \times \Z_2^\prime)$.

One possibility would be to add localized fermions at the fixed points,
analogous to the twisted sectors of string compactifications. However, 
as we have seen in section 2, a bulk fermion produces only half of the 
anomaly of a Weyl fermion at each fixed point. Therefore, this
possibility may be generically cumbersome to realize without the 
guidance of an underlying string theory.

Another possibility would be to
implement an anomaly cancellation mechanism of the Green--Schwarz
\cite{gs} or inflow \cite{ch} type. The former would lead to a
spontaneous breaking of the $U(1)$ gauge symmetry \cite{gs4}. For the
latter, we must cope with the fact that a 5-D Chern--Simons term 
$\epsilon^{MNOPQ} A_M F_{NO} F_{PQ}$ 
(see \cite{Arkani-Hamed:2001tb}) cannot 
be added to the bulk Lagrangian, because it is not invariant under the
two orbifold projections. However, we can imagine more general
possibilities.  For example~\footnote{We thank R. Rattazzi for having
suggested this possibility to us.}, we could introduce a bosonic field
$\chi$ with $(-,-)$ periodicities and try to introduce the
Chern--Simons term in combination with $\chi$. The field $\chi$ should
then dynamically get a vacuum expectation value with a non-trivial
$y$-profile 
\begin{equation}
  \label{eq:epsi}
  \langle \chi(y) \rangle \propto \left\{
    \begin{array}{cc}
      +1\,, & 0<y<\frac{\pi R}{2} \qquad (\mathrm{mod.}\; \pi R) \\
      -1\,, & \frac{\pi R}{2}<y< \pi R \qquad (\mathrm{mod.}\; \pi R)
    \end{array}
    \right.
\end{equation}
(breaking spontaneously the $\Z_2\times \Z_2^\prime$
discrete symmetry), thereby generating a sort of magnetic charge
for the fixed points and leaving only very massive fluctuations. 
The resulting Chern--Simons term could then cancel, for an appropriate 
value of the coefficient, the one-loop anomaly.
Notice that this mechanism can work only when the integrated anomaly 
vanishes.

It is interesting to observe that the presence and the structure of the
anomalies that we have found in the $S^1/(\Z_2 \times \Z_2^\prime)$ orbifold 
could have been anticipated~\footnote{We are grateful to A. Uranga for
discussions on this argument.} by analysing the intermediate models on 
$S^1/\Z_2$ or $S^1/\Z_2^\prime$. Indeed, the $S^1/\Z_2$ and $S^1/\Z_2^\prime$
models have anomalies given by eq.~(\ref{prevres}) and localized at the two 
$\Z_2$ and $\Z_2^\prime$ fixed points respectively, but with opposite sign, 
reflecting the difference between the $\Z_2$ and $\Z_2^\prime$ actions on 
the fermions.

For supersymmetric models, all the above considerations apply, but
supersymmetry poses further constraints.  It seems then quite
difficult to get a consistent SUSY field theory on the 
$S^1/(\Z_2 \times \Z_2^\prime)$ orbifold with a single
bulk Higgs hypermultiplet. On the contrary, the addition of a second Higgs
hypermultiplet in the bulk, as in \cite{Pomarol}, would cancel at the 
same time the anomaly and the one-loop-induced FI term. 
It therefore seems that the necessity of having two Higgs
doublets in 4-D supersymmetric extensions of the SM persists also in 
these higher-dimensional constructions.

\section*{Acknowledgements}

We would like to thank J.-P.~Derendinger, F.~Feruglio, P.~Gambino,
G.~Isidori, A.~Pomarol, A.~Uranga and especially R.~Rattazzi for
interesting discussions. This work was partially supported by the RTN
European Programs ``Across the Energy Frontier'', contract
HPRN-CT-2000-00148, and ``The Quantum Structure of Space-Time'',
contract HPRN-CT-2000-00131.  M.S. and L.S. acknowledge partial
support by CERN.  C.A.S. and M.S. thank INFN and the Phys. Dept. of
Rome Univ. ``La Sapienza'', where part of this work was done.

\end{document}